\documentclass{aa}  
\usepackage{graphicx}
\usepackage[breaklinks,pdfnewwindow=true,colorlinks=true,citecolor=blue]{hyperref}
\usepackage{txfonts}
\usepackage{mathrsfs}
\usepackage{orcidlink}
\DeclareUnicodeCharacter{03B2}{$\beta$}
\newcommand{\orcid}[1]{\unskip\protect\href{https://orcid.org/#1}{\protect\includegraphics[width=8pt,clip]{logo_orcid}}}

\usepackage{color}
\usepackage{soul}
\usepackage{xcolor}

\newcommand{\dd}{\mathrm{d}}

\definecolor{orange}{rgb}{1,0.5,0}
\definecolor{sred}{rgb}{.5,0,0}

\begin{document}

   \title{Impact of cosmic-ray propagation on the chemistry and\\ ionisation fraction of dark clouds}
   \subtitle{}

   \author{G. Latrille
          \inst{1}\fnmsep\orcidlink{0009-0004-3920-2010} \thanks{Corresponding author: gonzajaque2016@udec.cl}
          \and
          A. Lupi
          \inst{2,3}\orcidlink{0000-0001-6106-7821}
          \and
          S. Bovino\inst{1,4,5}\orcidlink{0000-0003-2814-6688}
          \and
          T. Grassi\inst{6}\orcidlink{0000-0002-3019-1077}
          \and 
          G. Sabatini\inst{5}\orcidlink{0000-0002-6428-9806}
          \and 
          M. Padovani\inst{5}\orcidlink{0000-0003-2303-0096}
          }

   \institute{
   Departamento de Astronomía, Facultad de Ciencias Físicas y Matemáticas, Universidad de Concepción, Av. Esteban Iturra s/n Barrio Universitario, Casilla 160, Concepción, Chile 
   \and
   Como Lake Center for Astrophysics, DiSAT, Universit\`a degli Studi dell'Insubria, via Valleggio 11, I-22100, Como, Italy 
   \and
   INAF, Osservatorio Astronomico di Bologna, Via Gobetti 93/3, I-40129 Bologna, Italy
   \and
   Chemistry Department, Sapienza University of Rome, P.le A. Moro, 00185 Rome, Italy
      \and
   INAF - Osservatorio Astrofisico di Arcetri, Largo E. Fermi 5, 50125 Firenze, Italy
   \and
   Max-Planck-Institut für Extraterrestrische Physik, Giessenbachstr. 1, D-85749 Garching bei München, Germany
 }

   \date{Received XXXX / Accepted ZZZZ}

  \abstract
   {}
   {A proper modelling of the cosmic-ray ionisation rate within gas clouds is crucial to describe their chemical evolution accurately. However, this modelling is computationally demanding because it requires the propagation of cosmic rays throughout the cloud over time. We present a more efficient approach that simultaneously guarantees a reliable estimate of the cosmic-ray impact on the chemistry of prestellar cores.
  }
   {We introduce a
   numerical framework that mimics the cosmic-ray propagation within gas clouds and applies it to magnetohydrodynamic simulations performed with the code \textsc{gizmo}. It simulates the cosmic-ray attenuation by computing the effective column density of H$_2$ that is traversed, which is estimated using the same kernel weighting approach as employed in the simulation. The obtained cosmic-ray ionisation rate is then used in post-processing to study the chemical evolution of the clouds.
   }
   {We found that cosmic-ray propagation affects deuterated and non-deuterated species significantly and that it depends on the assumed cosmic-ray spectrum. We explored correlations between the electron abundance, the cosmic-ray ionisation rate, and the abundance of the most relevant ions (HCO$^+$, N$_2$H$^+$, DCO$^+$, N$_2$D$^+$, and o-H$_2$D$^+$), with the purpose of finding simple expressions that link them. We provide an analytical formula to estimate the ionisation fraction, $X(\mathrm{e^-})$, from observable tracers and applied it to existing observations of high-mass clumps. We obtained values of about 10$^{-8}$, which is in line with previous works and with expectations for dense clouds. We also provide a linear fit to calculate the cosmic-ray ionisation rate from the local H$_2$ density, which is to be employed in three-dimensional simulations that do not include cosmic-ray propagation.}
   {}


   \maketitle

\section{Introduction}
    Dark clouds are commonly characterised by extreme conditions, such as temperatures below $20~\mathrm{K}$ and high column densities that span $10^{23} - 10^{25}~ \mathrm{cm}^{-2}$ \citep{bergin2007}. The dense material is sufficient to completely shield the cloud from incoming photo-radiation, which leaves only cosmic rays (CRs) as a potential source of ionisation in the inner regions of the clouds. 
    Cosmic rays are non-thermal charged particles whose origin depends on their energies, which extend from local sources such as stars ($10^9 - 10^{10}~\mathrm{eV}$), galactic sources such as supernova remnants ($10^{10} - 10^{15}~\mathrm{eV}$), and extra-galactic sources such as active galactic nuclei \citep[$ \mathrm{E} >10^{15}~\mathrm{eV}$; see][as a review]{padovani2020}.
    
    Within dark clouds, hydrogen is mainly in its molecular form, H$_2$, and it represents the major constituent, followed by CO and then other species. Because H$_2$ dominates the other species, CRs are expected to have a stronger impact on it and first form H$_2^+$ and subsequently H$_3 ^+$. After H$_3^+$ is formed, the path to molecular formation is paved. It leads to the production of other molecules such as the OH$^+$ and HCO$^+$ ions \citep{dalgarno2006,indriolo2013}, the molecular form of atomic species such as O$_2$ and N$_2$ \citep{indriolo2013}, complex neutrals, H$_2$O and NH$_3$ \citep{gaches2022}, and deuterated species such as H$_2$D$^+$ \citep{caselli2019}.

    The latter is formed through the reaction of HD with H$_3^+$ \citep[e.g.][]{dalgarno1984}. When it is formed, the path to deuteration begins. Because H$_2$D$^+$ donates its deuterium atom to abundant neutral species such as CO, DCO$^+$ is produced. At the same time, CO forms HCO$^+$ via interaction with H$_3^+$, 
    \begin{align}
        \mathrm{CO} + \mathrm{o\text{-}H}_2\mathrm{D}^+ \longrightarrow \mathrm{DCO}^+ + \mathrm{H}_2 ,\label{eq:dco+}\\
        \mathrm{CO} + \mathrm{H}_3^+ \longrightarrow \mathrm{HCO}^+ + \mathrm{H}_2 . \label{eq:hco+}
    \end{align}
    The formation of N$_2$H$^+$ is similar to that of HCO$^+$,
    \begin{equation}
        \mathrm{N}_2 + \mathrm{H}_3^+ \longrightarrow \mathrm{N}_2\mathrm{H}^+ + \mathrm{H}_2 . \label{eq:n2h+}
    \end{equation}
    
    Ionisation occurs through collisions between CRs and neutral molecules (H$_2$, N$_2$, and CO). This drives fast ion-neutral reactions in the environment. 
    The number of ionisations of H$_2$ molecules (or H atoms) per unit time, referred to as the CR ionisation rate, therefore plays a fundamental role in determining the chemical state of star-forming regions. Throughout this paper, we indicate the ionisation rate for molecular hydrogen as $\zeta_2$. 
    
    Several studies have attempted to determine the CR ionisation rate from observational tracers in different environments so far, including diffuse medium \citep{shaw2008, indriolo2012b}, high-mass star-forming regions \citep{sabatini2020, sabatini2023}, circumstellar discs \citep{ceccarelli2014}, and low-mass dense cores \citep{caselli1998, redaelli2021, bialy2022}. To do this, they relied on relations between the species abundances and $\zeta_2$ \citep{caselli1998, indriolo2007, indriolo2012a, indriolo2012b, bovino2020,socci2024}, or they explored the chemical evolution with low-dimensional (0D or 1D) models \citep{shaw2008, shingledecker2016, neufeld2017,  redaelli2021}. From a theoretical point of view, it is highly computationally expensive to propagate CRs in simulations alongside hydrodynamics because the particle velocities are high (close to the speed of light). This reduces the simulation time-step, the extended energy spectrum, and several uncertainties in the CR propagation process itself. Some studies have attempted to self-consistently couple CRs with magnetohydrodynamics only recently \citep[MHD;][]{chan2019, winner2019, bustard2021, ogrodnik2021, thomas2021}, but with several limitations, and never alongside a proper non-equilibrium chemical evolution. 
    \newline \indent We explore the effect of a consistent CR propagation on the chemistry of prestellar cores simulated with the magnetohydrodynamics code \textsc{gizmo} \citep{hopkins2015,hopkins2016},
    which was previously reported by \citet{bovino2019}. 
    The paper is organised as follows: In Section~\ref{sec:framework} we describe the numerical framework, that is, the CR propagator and the coupling of the CR ionisation rate to the chemical post-processing. In Section~\ref{sec:results} we discuss our main results, and in Section~\ref{sec:conclusions} we draw our conclusions.

\section{Numerical framework}
\label{sec:framework}
Our numerical framework is composed of two parts, a CR propagation code that we describe in Section~\ref{sec:crscheme}, and a chemical post-processing tool that we present in Section~\ref{sec:chemistry}. The tool uses the output of the CR propagation code as input.
\subsection{CR propagation scheme}\label{sec:crscheme}

\begin{figure*}[h!]
    \centering
    \includegraphics[width=0.9\textwidth]{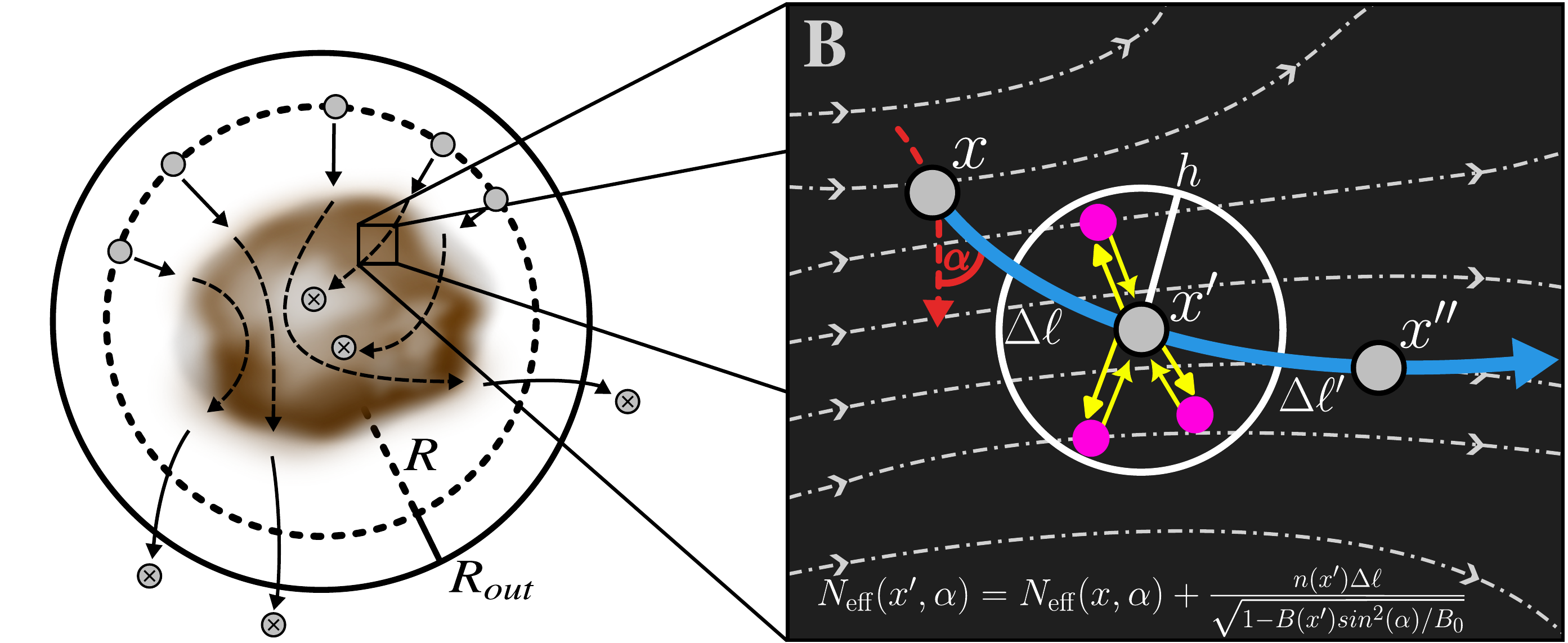}
    \caption{ Sketch of the CR propagation scheme based on the structure given in Section \ref{sec:crscheme}. The left panel describes the TP propagation through the domain. TPs (grey dots) are created from a spherical surface of radius $R$ (dashed black circle) with radial inward trajectories (step i). TPs are then deactivated when the column density that is traversed exceeds 10$^{29}$~cm$^{-2}$ or when they leave the outer spherical surface at a distance $R_\mathrm{out}$ (solid black circle), as shown by the crossed grey dots. The right panel instead highlights the propagation of a single TP particle. The trajectory is shown by the curved solid blue arrow, and the magnetic field lines ($B$) are reported as dash-dotted grey lines. The pitch angle ($\alpha$) between the TP velocity and the local magnetic field is shown as the dashed red line. The displacement of the TP between the previous ($\mathbf{x}$) and the current ($\mathbf{x}'$) location is defined by the step size $\Delta\ell$ (step iii). To compute the effective column density of the TP, the average local density ($n$($\mathbf{x}'$)) and local magnetic field ($B$($\mathbf{x}'$)) were estimated from gas particles (pink dots) within the kernel size $h$ (thin yellow arrows; step ii). Then, its value was used to update the CR ionisation rate ($\zeta_2$; step iv), which was finally deposited at the location of the original gas particle (thick yellow arrows; step v).}
    \label{fig:CRpropag}
\end{figure*}

    The propagation scheme we considered aimed to mimic a continuous CR flux in which CRs are modelled as tracer particles (TPs) that propagate throughout the system (e.g. a molecular cloud). Because CRs are charged particles, they propagate along paths defined by the magnetic field lines. As CRs propagate through the cloud, they lose energy because of collisions with molecular gas. This results in an effective attenuation of the flux deeper in the cloud at increasingly higher column densities \citep{padovani2024}.

    The structure of the propagation scheme is as follows:
\begin{enumerate}[(i)]
    \item TP generation: A specific number of TPs are created (determined by $2^p$, with $p$ set before the run) at a distance $R$ from the cloud, and they are distributed randomly over a spherical surface. The initial direction of the TPs along the magnetic field lines is chosen to be towards the centre of the cloud, and all are assumed to start from a uniform isotropic distribution of pitch angles ($\alpha _0$) over a hemisphere, where the pitch angle distribution ranges from 0 to $\pi/2$, sampled using $N_\alpha$ bins. For each initial value of the pitch angle, TPs are assigned an initial column density 
    \begin{equation}
        N_{\rm eff}(\mathbf{x}_0,\alpha_0) = \frac{2\times 10^{21}\rm\, cm^{-2}}{\cos{\alpha_0}},
    \end{equation}
    where $N_{\rm eff}$ stands for the effective column density that is traversed by the TP, $2\times 10^{21}$~cm$^{-2}$ is the column density value at which hydrogen is mainly in its molecular form \citep[][]{snow2006}, and $\mathbf{x}_0$ is the initial position of the TP.
    \item Gas particle interpolation: In order to compute the column density that is traversed by CRs and then the CR ionisation rate, local gas properties at the position of the TPs are needed, in particular, the H$_2$ number density $n(\mathrm{H}_2)$ and the average magnetic field $B$, the latter setting the propagation direction. Because all these physical quantities vary throughout the cloud, we estimated them by means of kernel weighting, analogous to what is done for hydrodynamics \citep[see][]{hopkins2015}. In particular, at the location $\mathbf{x}_k$ of the $k$-th TP, we determined the kernel size $h_k$ that encompasses an effective number of gas-neighbours $N_{\rm ngb}=32$, defined by 
    \begin{equation} 
        N_{\rm ngb} = \frac{4\pi}{3}h_k^3 \sum _j W(\mathbf{x}_j - \mathbf{x}_k, h_k), \label{eq:smooth}
    \end{equation}
    where the sum is over all neighbours within $h_k$, $\mathbf{x}_j$ is the location of the $j$th neighbour, and the kernel function $W(\mathbf{x}_j-\mathbf{x}_k,h_k)$ is defined by the standard cubic spline 
    \begin{eqnarray}
     \bar{W}(q,h_k) = \frac{8}{\pi h_k^3} 
    \begin{cases}
        1-6q^2 + 6q^3, & 0 \leq  q \leq \frac{1}{2}, \\
        2(1-q)^3, & \frac{1}{2} < q \leq 1, \\
        0, & q > 1, 
    \end{cases}
        \label{eq:spline}
    \end{eqnarray}
    with $q = |\mathbf{x}_j-\mathbf{x}_k|/h_k$ the normalised distance.

    The neighbour search is repeated at each step of the integration for all TPs and thus represents the highest computational cost of the scheme because Eq.~\eqref{eq:smooth} must be solved implicitly. To limit the computational cost and rapidly converge to the desired solution, we performed this operation using a Newton-Raphson iteration\footnote{We verified that convergence was typically reached in up to three iterations.} \citep{springel2005,hopkins2015}.
    After the kernel size was determined, a last loop on the neighbours was performed to compute the smoothed magnetic field and the gas density, which were then used to update the column density.
    
    \item TPs propagation: The propagation of TPs is computed by means of a fourth-order Runge-Kutta integrator, where the step size for the $k$th TP is determined as $\Delta \ell_k=0.1\Delta x_k$, with $\Delta x_k = (4\pi/3)^{1/3}N_{\rm ngb}^{-1/3}h_k$ the grid-equivalent inter-particle spacing, and motion occurs along the direction of the smoothed magnetic field. We note that the interpolated quantities were recomputed at every sub-step of the propagation by repeating (ii).
        
    \item Column density and CR ionisation rate update:
    Using the new location of the TP $\mathbf{x}'$ and the new interpolated quantities $n(\mathbf{x}')$ and $B(\mathbf{x}')$, we updated the column density that is traversed by CRs as 
     \begin{equation} 
         N_{\mathrm{eff}}(\mathbf{x}',\alpha_0) = N_{\mathrm{eff}}(\mathbf{x,\alpha_0})+\frac{n(\mathbf{x}')\Delta \ell_k}{\sqrt{1-B_k(\mathbf{x}')\sin^2 (\alpha_0)/B_{k,0}}},
         \label{eq:Neff}
     \end{equation} 
     and the corresponding CR ionisation rate was re-estimated by integrating with the trapezoidal rule over all the pitch angles as
    \begin{equation}
        \zeta _2  (\mathbf{x}') = \delta(\mathbf{x}') \int _0 ^{\pi/2} f\left[N\mathrm{_{eff} (\mathbf{x}', \alpha)} \right]\sin{\alpha}\,\dd\alpha, \label{eq:CR_estimate}
    \end{equation}
    where $\delta(\mathbf{x}')=\frac{B(x')}{B_0}$ accounts for the increase in $\zeta_2$ due to the growth in the magnetic flux, with $B(\mathbf{x}')$ the magnetic field at the current position and $B_0$ the field at the TP injection, $\sin \alpha$ accounts for the spherical distribution of the pitch angles \citep[see Eq.~(6) in][]{padovani2013} at the TP injection, and the $f$ function converts $N_\mathrm{eff}$ into the corresponding $\zeta _2$ according to the relations presented in 
    Fig.~F.1 of \cite{padovani2018}, where the three curves represent different levels of $\zeta_2$ and account for the ionisation of H$_2$ by protons and primary and secondary electrons. The exact conversion defined by $f$ is arbitrary and was chosen at the beginning of the calculation. We considered model $\mathscr{H}$ (the average CR ionisation rate on diffuse media\footnote{Note that recently the estimates of $\zeta_2$ for diffuse clouds have been revised in \citet{obolentseva2024}. This model would then represent an upper limit to diffuse clouds estimates.}) and model $\mathscr{L}$ in particular, which comes from the data of the two Voyager spacecraft \citep[see][for more details]{padovani2022}. 

    In our set-up, the argument of the square root in Eq.~\eqref{eq:Neff} could become negative or equal to zero. When this occurs, the gyrocenter of the particle is mirrored backwards \citep[e.g.,][]{silsbee2018}. It is therefore possible to define $B_{\rm crit}=B_0/\sin^2(\alpha _0)$ as the upper limit of the non-mirroring particles. The propagation scheme includes this effect by assuming that the reflected particles above $B_{\rm crit}$ are neglected in the evaluation of Eq.~\eqref{eq:CR_estimate}.

    \item CR ionisation rate deposition: At the end of each iteration of the propagation algorithm, the $\zeta_2$ value for each $k$th TP is deposited on the $j$th gas particles that contributed to its calculation, again weighted by the smoothing kernel. Because the same gas particles can contribute to the calculation of $\zeta_2$ at different iterations, the actual value of $\zeta_2$ for each gas particle was computed only at the end of the entire procedure as
    \begin{equation}
        \zeta _{2} ^{j} = \frac{\sum _{k=1} ^{N_p} W(\mathbf{x}_j-\mathbf{x}_k,h_k) \zeta_2 (\mathbf{x}_k)}{\sum _{k=1} ^{N_p} W(\mathbf{x}_j-\mathbf{x}_k,h_k)}.
    \label{eq:deposit}
    \end{equation}
\end{enumerate}

\begin{figure}
    \centering
    \includegraphics[width=\columnwidth]{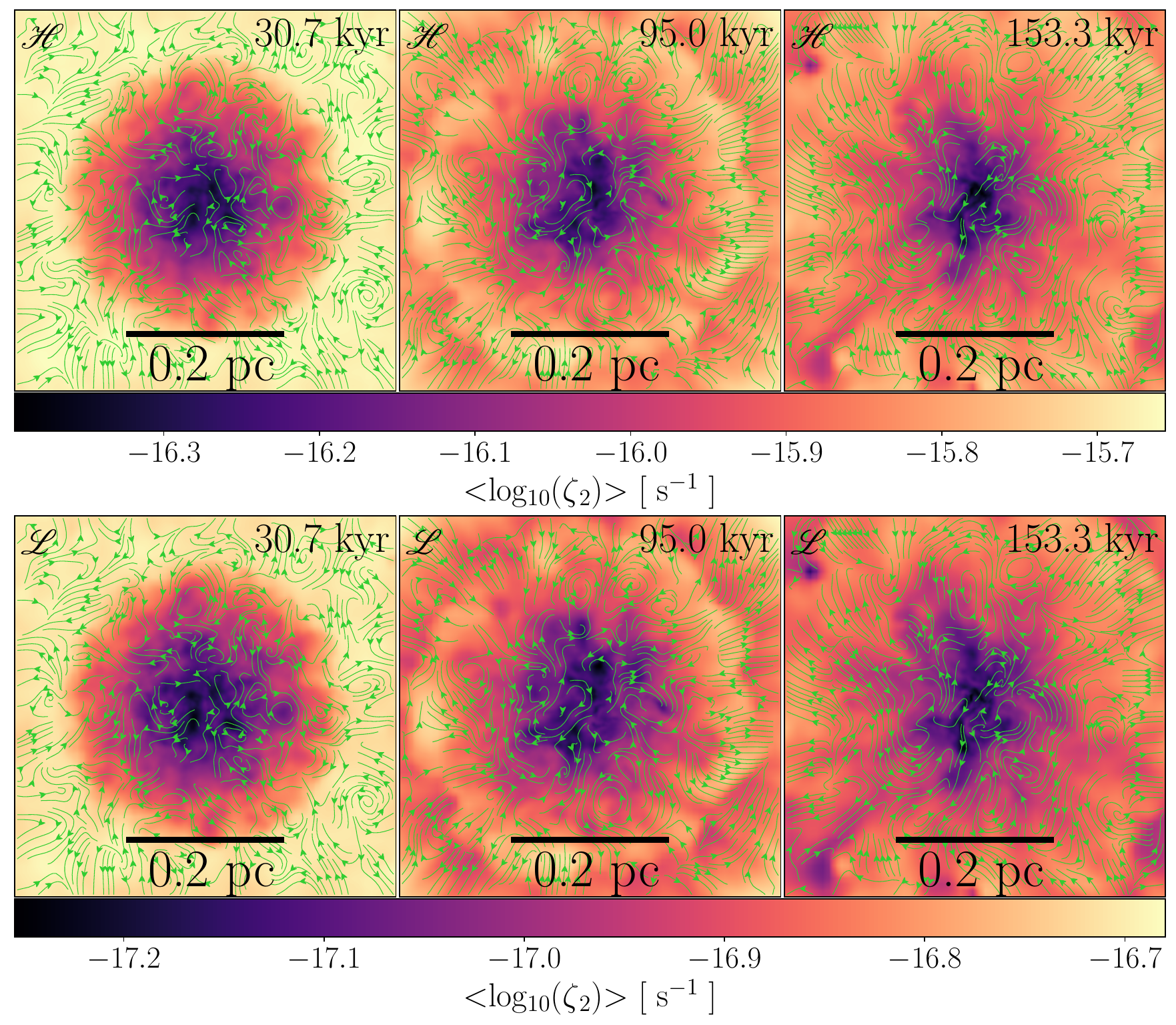}
    \caption{Time evolution of the density-weighted $\zeta _2$ for models $\mathscr{H}$ (top panels) and $\mathscr{L}$ (bottom panels) in a sphere of 0.5~pc radius, inside which CRs are propagated. The green arrows correspond to the  x-y projection of the magnetic field lines. The two different models yield a difference of one order of magnitude within the high-density region. }
    \label{fig:Crir_Three}
\end{figure}
    Steps (ii) to (iv) were repeated over all active TPs, where a TP was considered active until one of the following two conditions was fulfilled: a) the CR flux becomes negligible, which in our scheme corresponds to reaching a traversed column density $N_{\mathrm{eff}}$ > 10$^{29}$~cm$^{-2}$, or b) TPs leave the spherical region around the cloud, defined by an outer boundary spherical surface, which in our case was located at $R_{\rm out}=1.02\max_j\{r_j\}$, where $r_j$ is the distance of the $j$th gas particle from the centre of the cloud. The calculation ended when all the TPs became inactive, namely, they completed their propagation throughout the cloud.
    
    The final output of our propagation algorithm, a sketch of which is shown in Fig.~\ref{fig:CRpropag}, consisted of a 3D distribution of $\zeta_2$ generated from the magnetised cloud at a specific time during its evolution. By taking different time steps from the simulation, it is then possible to obtain the evolution of the $\zeta_2$ distribution over time as the gas evolves and to use it as an input to the chemical post-processing, as explained in the next section.

\subsection{Chemical post-processing}
\label{sec:chemistry}
    The output of the CR propagator can be used to chemically evolve the cloud with time-dependent chemistry over the timescale of the simulation. The calculations were performed via the post-processing tool developed by \cite{ferrada-chamorro2021}, which was coupled to the thermo-chemistry library {\sc krome} \citep{grassi2014}. The chemical network we employed includes 134 species and 4616 reactions \citep[for more details, see the `M1' case and Section 2 of][]{bovino2019} and was evolved throughout the entire history of the cloud. We obtained the $\zeta _2$ distribution at each timestep, considering models $\mathscr{H}$, $\mathscr{L}$, and an additional reference case ($\mathscr{C}$), where the CR ionisation rate was kept constant to a typical value of $\zeta_2 = 2.5\times 10^{-17}~\mathrm{s^{-1}}$ (hereafter $\zeta _c$). We used $\zeta_c$ to compare the impact of a consistent propagation of CRs on several species (HCO$^+$, N$_2$H$^+$, DCO$^+$, N$_2$D$^+$,  and o-H$_2$D$^+$) that are commonly used as tracers for $\zeta_2$.
    We first determined the value of $\zeta_2$ every three snapshots of the simulation, corresponding to a time interval of 1.5 kyr. This was done to decrease the computational costs. Following \cite{ferrada-chamorro2021}, we only considered 10\% of all gas particles and interpolated the chemical abundances for the remaining ones. The subset particles were randomly selected according to the procedure and the convergence tests reported by \cite{ferrada-chamorro2021}, which guarantee a proper ratio of the core and background particles.

\section{Results} 
\label{sec:results}
\subsection{Distribution of the CR ionisation rate}
    
Before investigating the chemical evolution in the core, we analysed the distribution of $\zeta_2$ resulting from our propagator. The code was applied to each snapshot of the simulation within a spherical region of radius $0.5~\mathrm{pc}$ centred on the highest-density region of the cloud.
    
\begin{figure}[h!]
    \centering
    \includegraphics[width=1\columnwidth]{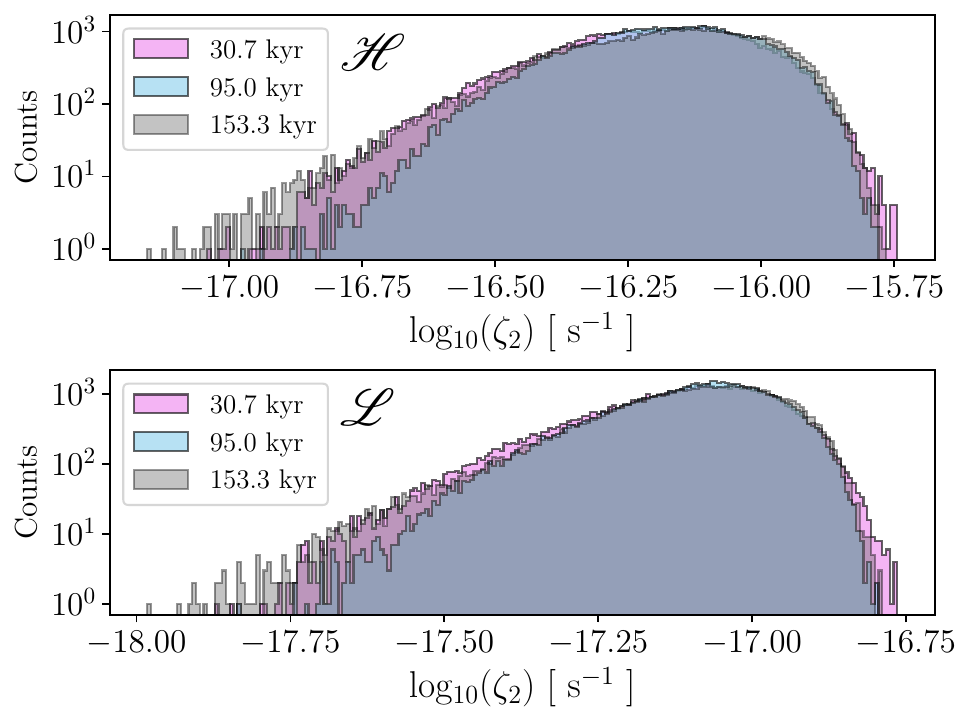}
    \caption{Time evolution of the distribution of $\zeta _2$ for the gas particles in the simulation for models $\mathscr{H}$ (top) and $\mathscr{L}$ (bottom). The distributions are determined from a box of 0.2$\times$0.2$\times$0.2~pc$^3$, which only includes particles from the simulated cloud, i.e., with identical mass. The rightmost edge of the distribution shrinks over time as the gas particles within the box become denser.
    The low-end tail of the distribution expands towards lower ionisation rates, which is consistent with the evolution of the gas density.}
    \label{fig:Time_evol}
\end{figure}
    
    Figure~\ref{fig:Crir_Three} shows density-weighted maps of $\zeta_2$ for $\mathscr{H}$ (top panels) and $\mathscr{L}$ (bottom panels) at three different times: 30.7, 95.0, and 153.3~kyr. The x-y projection of the magnetic field lines is overplotted as green arrows. $\zeta _2$ differs by approximately one order of magnitude between the two models and spans a range of 0.6 dex. The interval below 10$^{-16.2}$~s$^{-1}$ ($\mathscr{H}$) and 10$^{-17.1}$~s$^{-1}$ ($\mathscr{L}$) covers the high-density region, as expected by considering the CR attenuation, and the highest CR ionisation rate is found in the background region where CRs are almost unattenuated. Strong twisting and bending of the magnetic field lines is also observed, especially in the highest-density regions, where the line density (hence the magnetic field intensity) also increases. This complex pattern results in a significant increase in the effective column density that is traversed by CRs, and hence, in a stronger attenuation.
    
    The time evolution of the $\zeta _2$ distribution for the two models in a 0.2$\times$0.2$\times$0.2~pc$^3$ box centred at the peak of the high-density region is shown in Fig.~\ref{fig:Time_evol}. 
    Both distributions show that regardless of the initial $\zeta _2$ computed at injection, the attenuation reduces the initial ionisation rate by more than one order of magnitude.
    
     In the intervals 10$^{-16.4}$ to 10$^{-15.8}$~s$^{-1}$ ($\mathscr{H}$) and 10$^{-17.3}$ to 10$^{-16.7}$~s$^{-1}$ ($\mathscr{L}$), the $\zeta _2$  distribution does not vary over time, but for the low- and high-end tails. In particular, the low-$\zeta_2$ tail, which is associated with the innermost region of the cloud (namely only about 1\% of the total mass) simply fluctuates over time without any clear evolutionary trend. When the lowest ionisation rates per distribution, $\sim10^{-17.1}$ ($\mathscr{H}$) and $\sim10^{-18}$~s$^{-1}$ ($\mathscr{L}$), are compared with Fig.~F.1 of \cite{padovani2018}, these ionisation rate values correspond to column densities of $\sim10^{25-26}$ cm$^{-2}$. In our run, this is the effect of the pitch angles that were not considered in Eq.~\eqref{eq:CR_estimate}, and it is related to the mirroring effect and not to a high column density accumulated by TPs.
     
\subsection{$\zeta _2$ dependence on local properties}

\begin{figure}[h!]
    \centering
    \includegraphics[scale=0.5]{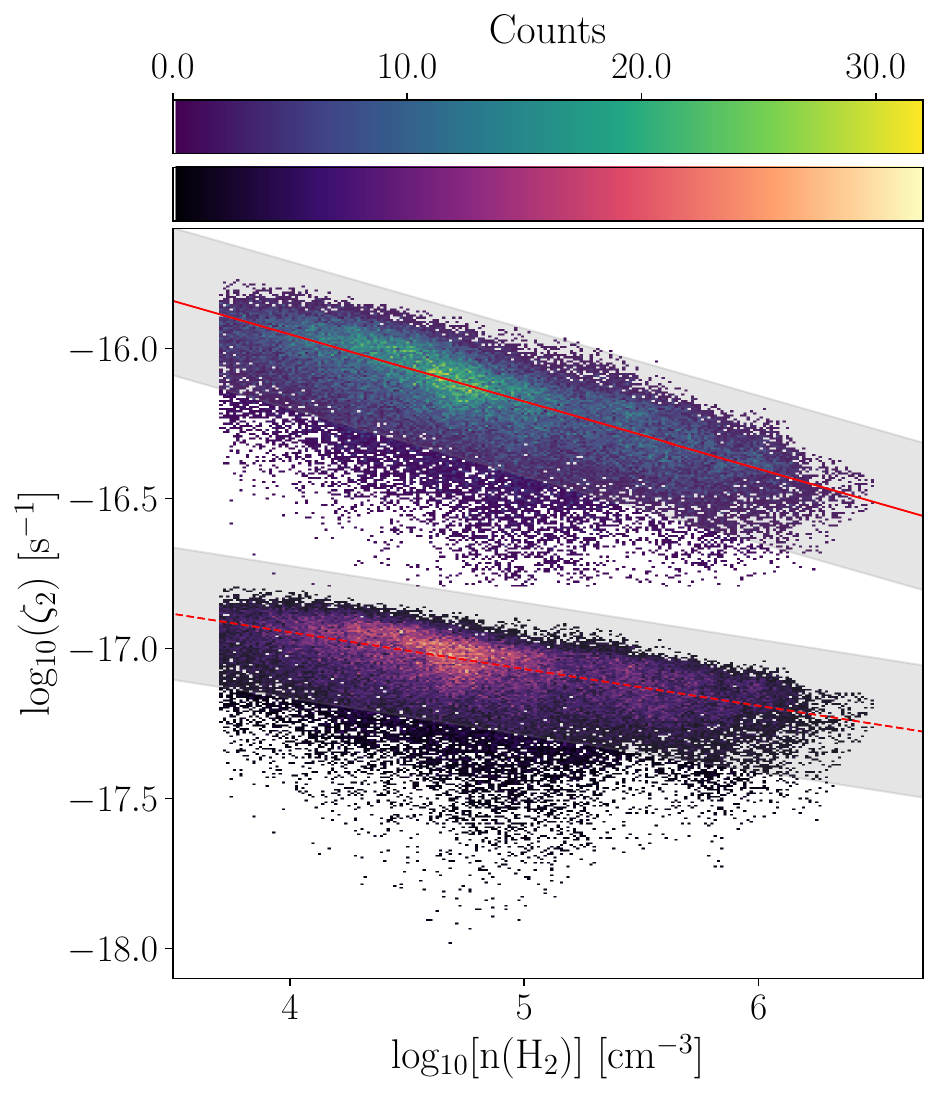}
    \caption{Correlation between $\mathscr{\zeta}_2$ and n(H$_2$) at 153.3~kyr. The viridis and magma color maps represent the $\mathscr{H}$ and $\mathscr{L}$ models, respectively. The solid and dashed red lines correspond to our best-fit relations, and the shaded grey areas show the intrinsic scatter discussed in Section 3.2.}
    \label{fig:local_zeta_nH2}
\end{figure}

Because the self-consistent propagation of CRs in numerical simulations is expensive, effective models that relate the gas density to $\zeta_2$ are often employed. With our framework, we have the unique ability to determine more accurate semi-analytic prescriptions and how these prescriptions might depend on the CR model that is considered. The attenuation of $\mathscr{\zeta}_2$ is expected to significantly depend on the gas density, under the assumption that high volume densities correspond higher column densities \citep{sabatini2023, socci2024}. In Fig.~\ref{fig:local_zeta_nH2} we explore this dependence by showing the correlation between the CR ionisation rate and the number density of H$_2$, where the viridis histogram corresponds to the $\mathscr{H}$ model and the magma histogram to the $\mathscr{L}$ model. The trend is clearly negative in both models, which we fit with power laws. The power law is shown as solid and dashed red lines and given by the following expressions: 
    \begin{align}
       \log_{10}\mathscr{\zeta}_2 ^{\mathscr{H}} &= -0.22\log _{10} n(\rm{H}_2) -15.06, \label{eq:local_zeta_nH2_1}\\
       \log_{10}\mathscr{\zeta}_2 ^{\mathscr{L}} &= -0.12\log _{10} n(\rm{H}_2) - 16.45,
       \label{eq:local_zeta_nH2_2}
    \end{align}
    for models $\mathscr{H}$ and $\mathscr{L}$ respectively. 
    Due to the 3D structure of the system, the different lines of sight result in a slightly different CR ionisation rate, with a scatter of $\sigma ^\mathscr{H}=0.25$ and $\sigma ^\mathscr{L}=0.22$ for the two models (shown as shaded grey areas), calculated as the root mean square (RMS) between the actual $\mathscr{\zeta}_2$ and our best fits.

\subsection{Ion chemistry and physical correlations}
\begin{figure}[h!]
    \centering
    \includegraphics[width=\columnwidth]{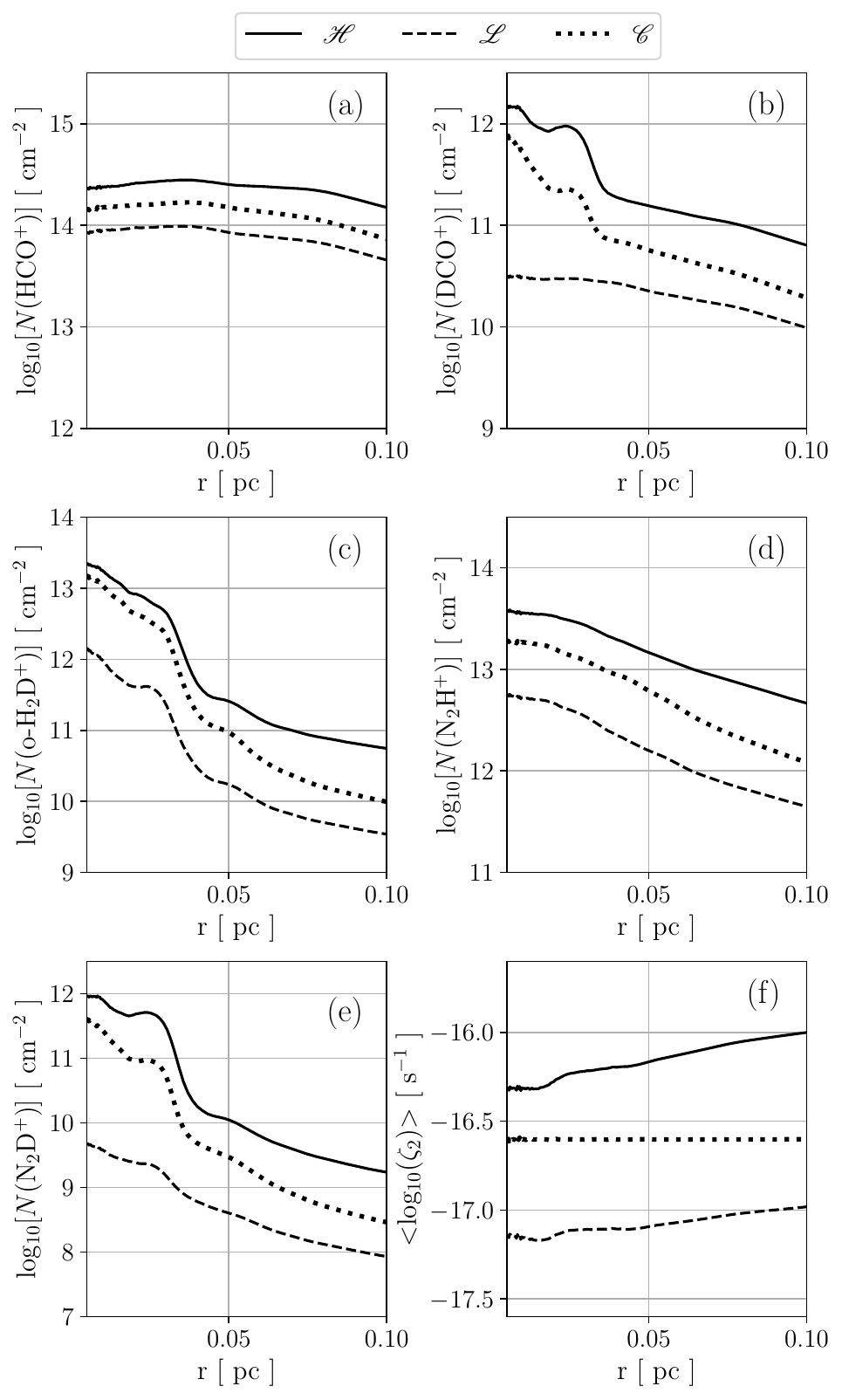}
    \caption{Radial profiles of the column densities of HCO$^+$ (a), DCO$^+$ (b), o-H$_2$D$^+$ (c), N$_2$H$^+$ (d), and N$_2$D$^+$ (e) ions and the $\zeta _2$ distribution (f) at 153.3~kyr. The $\zeta_2$ radial profile is calculated in circular annuli from the average along the line of sight (as shown in Fig.~\ref{fig:Crir_Three}). The CR ionisation models $\mathscr{H}$, $\mathscr{L}$, and $\mathscr{C}$ are represented by solid, dashed, and dotted black lines, respectively.}
    \label{fig:radial}
\end{figure}

    We present in Fig.~\ref{fig:radial} the radial profiles obtained from column density maps of the ions HCO$^+$, N$_2$H$^+$, DCO$^+$, N$_2$D$^+$, and o-H$_2$D$^+$, together with $\zeta_2$ for the inner 0.1 pc of the simulated prestellar core. The profiles display the $\mathscr{H}$ and $\mathscr{L}$ models and $\mathscr{C}$ at time 153.3~kyr. The column densities were calculated by integrating the gas cloud along a line of sight that was aligned with the $z$-axis over a depth of 0.2~pc. In order to exclude background gas outside the initial Bonnor-Ebert sphere representing our core, we removed gas particles with number densities below $\log[n/({\rm cm^{-3}})]<-3.78$, with $n$ calculated as $\rho(2.3m_{\rm p})^{-1}$ with $m_{\rm p}$  the proton mass. The radial profiles were then computed by averaging the integrated quantities in circular annuli. The radial profiles (panel f) clearly show the effect of the CR attenuation in the central part (highest density). This brings the ionisation rate from model $\mathscr{H}$ close to $\zeta_c$. As a consequence, for deuterated species in particular (panels b, c, and e), the column densities of the analysed tracers approach each other. The $\mathscr{L}$ model shows more marked differences, in particular, for deuterated species such as DCO$^+$ and N$_2$D$^+$. This is due to the efficiency of the formation channels. In particular, after checking the reaction fluxes in the different cases, we found that reactions that are directly dependent on $\zeta_2$ and on (multiple) deuterated species (e.g. H$_2$D$^+$, D$_2$H$^+$, and D$_2^+$) become much more relevant for high $\zeta_2$ values by boosting N$_2$D$^+$ and even more DCO$^+$, as a consequence of the CO + N$_2$D$^+ \rightarrow$ N$_2$ + DCO$^+$ reaction. Overall, the effect of the different CR ionisation rate spectra is highly non-linear because of the attenuation effects and the different chemical evolution of these species over time.

\begin{figure}
    \centering
    \includegraphics[width=8cm]{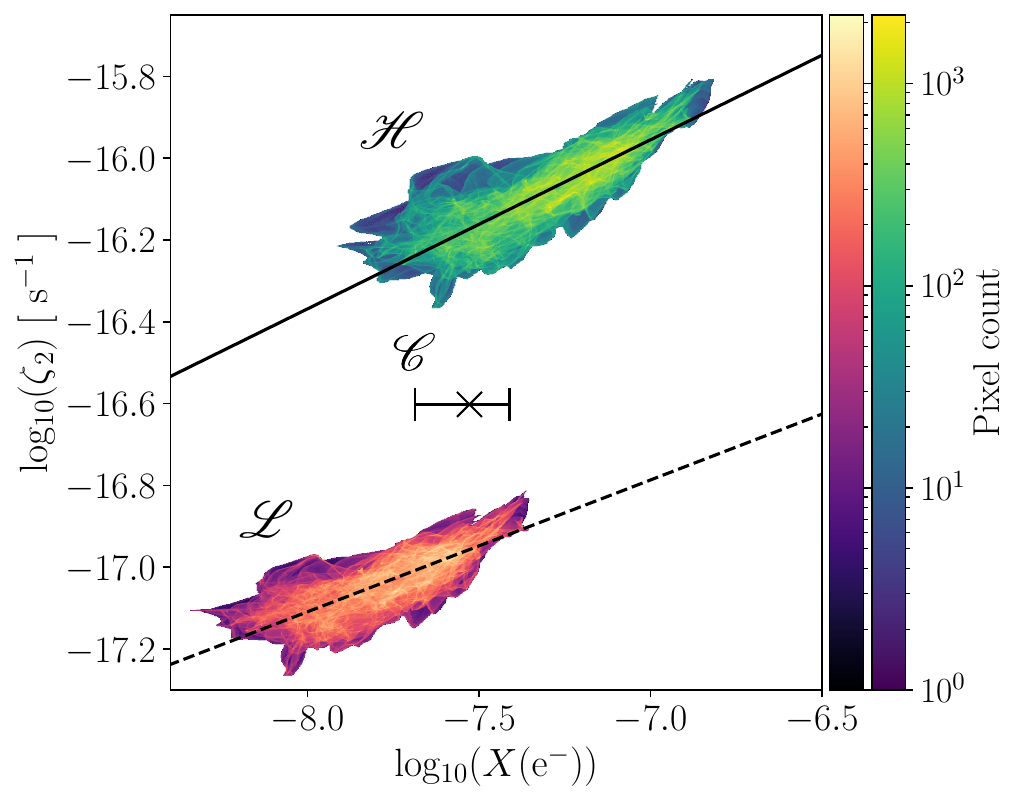}
    \caption{Correlation between $X$(e$^-$) and $\zeta _2$ obtained from the 2D maps (line-of-sight averages) for models $\mathscr{H}$ (outermost color scale) and $\mathscr{L}$ (innermost color scale), combining different time snapshots ($t=$ 30.7, 95.0, and 153.3~kyr). The solid and dashed lines correspond to power-law fits. The cross located at $\log_{10}(\zeta_c) = -16.6$ represents the median of the electron abundance ($X$(e$^-$)) for the $\mathscr{C}$ model, and the error bars are the 1$\sigma$ of the data.}
    \label{fig:ezeta}
\end{figure}

    We show in Fig.~\ref{fig:ezeta} the correlation between the abundance of electrons $X$(e$^-$) and $\zeta_2$ over time for the two models with variable CR ionisation rates.  The abundance was calculated from column densities obtained by integrating the gas cloud properties along the $z$-axis of the line of sight, as $X({\rm mol})=N({\rm mol})/N({\rm H}_2)$. The results for the $\mathscr{H}$ model are shown using the outermost color map, and those for model $\mathscr{L}$ are reported using the innermost color map.
    Overall, the two quantities show a clear linear correlation that remains stable over time. The distributions were fit with a power law of the form
    \begin{align}
        \log_{10}\zeta _2^\mathscr{H} &= 0.41\log_{10}{X({\rm e}^-)}-13.06, \\
        \log_{10}\zeta _2^\mathscr{L} &= 0.32\log_{10}{X({\rm e}^-)}-14.53.
    \end{align}

\begin{figure}
    \centering
    \includegraphics[width=8cm]{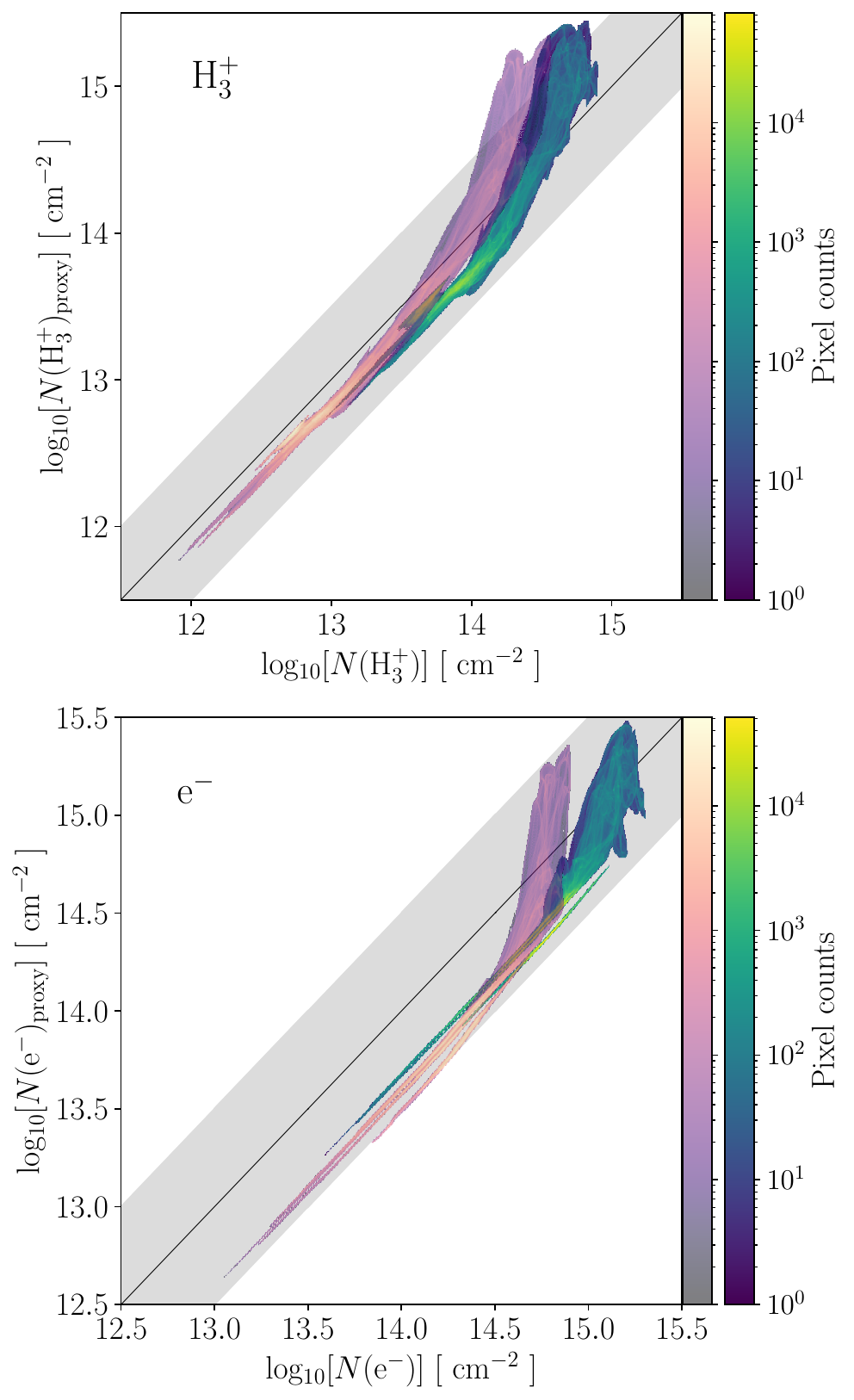}
    \caption{Pixel-by-pixel 2D histogram distributions for the $\mathscr{H}$ (viridis) and $\mathscr{L}$ (magma) model by combining different times ($t=$ 30.7, 95.0, and 153.3~kyr). The upper panel correlates the H$_3^+$ column density obtained from {\sc krome} and H$_3 ^+$ calculated as Eq.~\eqref{eq:h3p_proxy}. The bottom panel shows the correlation between the e$^-$ column density and its proxy calculated by Eq.~\eqref{eq:e_proxy}. The solid black line represents a 1:1 straight line and serves as an indicator for the lineal correlation formed by the 2D histogram distributions. The grey shadow indicates a region 
    spanning one order of magnitude (-0.5 and +0.5) around the 1:1 black line.}
    \label{fig:h3+_e-}
\end{figure}

    At first approximation, a linear correlation is also expected for the electron abundance (or column density) and the sum of the main positive ions (DCO$^+$, HCO$^+$, N$_2$H$^+$, N$_2$D$^+$, and H$_3^+$). Considering that H$_3^+$ is not observable in the dense and cold regions of molecular clouds, we replaced it with the proxy provided by \cite{bovino2020},
    \begin{align}
        N(\mathrm{H_3^+})_{\rm proxy} &= \frac{1}{3}\frac{N(\mathrm{o\text{-}H_2D^+})}{R_{\rm D}}, \label{eq:h3p_proxy} 
    \end{align}
    \noindent{where}
    \begin{align}
        R_{\rm D} &= \frac{N(\mathrm{DCO^+})}{N(\mathrm{HCO^+})},
    \end{align}
    and we tested the reliability of the following formula that can also be applied to observational data:
    \begin{align}\label{eq:e_proxy}
    \begin{split}
        N(\mathrm{e^-})_\mathrm{proxy} &= N(\mathrm{DCO^+}) + N(\mathrm{HCO^+}) + N(\mathrm{N_2D^+}) \\ 
        &+ N(\mathrm{N_2H^+}) + N(\mathrm{H_3^+})_\mathrm{proxy}. 
    \end{split}
    \end{align}
    We show in Fig.~\ref{fig:h3+_e-} the distributions of H$_3^+$ and e$^-$ against their proxies for models $\mathscr{H}$ and $\mathscr{L}$ averaged over time. The upper panel shows a very tight linear correlation between $N$(H$_3 ^+$) and $N$(H$_3 ^+$)$_{\mathrm{proxy}}$ up to $\sim10^{13}$-$10^{14}~\rm cm^{-2}$ for both models. A deviation then appears, while the relation still remains linear, but becomes steeper. This region corresponds to the central 10000 au, and the deviation is likely due to the neglected isotopologues in the estimation of H$_3^+$ \citep[see][for a discussion]{bovino2019}. A similar behavior is also observed for the electron column density (bottom panel of Fig.~\ref{fig:h3+_e-}). 
    
    Our proxy tends to underestimate the true electron abundance by  65\% overall. This was determined as the RMS error of the ratio of the distribution of $N({\rm e^-})_{\rm proxy}$ and the actual $N({\rm e^-})$ because only a limited number of ions was included in the formula and the approximations made. 
    The analytical formula reported in Eq.~\eqref{eq:e_proxy} was validated by employing a simulation of a typical prestellar core, specifically, on sub-parsec scales, but it also holds on larger scales and has a wider and more general applicability within the approximation we introduced so far. First,
    in the sum we neglected the contribution of ions that might become important (e.g. H$_3$O$^+$ among others) and the charged grains that might be relevant in star-forming regions, which particularly affects the total negative charge. In addition, when we probed H$_3^+$ through its first deuterated product (H$_2$D$^+$), we neglected other isotopologues that become important when the core central densities increase (see also \citealt{bovino2020,redaelli2024}). Finally, we worked with averaged quantities (column densities) to estimate local physical variables. For these reasons, the estimates of the ionisation fraction we provide represent a robust (within a factor of two) lower limit of the real value.\\

\subsection{Ionisation fraction in observed clumps}
\indent To show an application of the analytical formula presented in Eq.~\eqref{eq:e_proxy}, we employed recent available data for five high-mass clumps at different evolutionary stages: namely G13.18+0.06, G14.11-0.57, G14.63-0.58, G15.72-0.59, and G19.88-0.54. These sources were selected from the APEX\footnote{Atacama Pathfinder EXperiment 12 meter submillimeter telescope (APEX; \citealt{gusten2006}). APEX has been a collaboration between the Max Planck Institute for Radioastronomy, the European Southern Observatory, and the Onsala Space Observatory.} Telescope Large Area Survey of the Galaxy (ATLASGAL; \citealt{schuller2009}), where they have reliable estimates of kinematic distances, masses, luminosities, and dust temperatures \citep[e.g.][]{urquhart2022}, employed in this derivation.\\
\indent We used the data presented by \cite{sabatini2020, sabatini2024} for all tracers in Eq.~\eqref{eq:e_proxy}. \cite{sabatini2020} observed o-$\mathrm{H_2D^+}$ ($1_{10} - 1_{11}$), N$_2$H$^+$ (4-3), DCO$^+$ (1-0), and H$^{13}$CO$^+$ (1-0) with the APEX and IRAM-30m telescopes. Additional N$_2$D$^+$ (3-2) data were obtained with the nFLASH receiver mounted at APEX through two projects (project-IDs: C-0105.F-9715C-2020 and C-0107.F-9711-2021; PIs: S. Bovino and G. Sabatini, respectively). We refer to \cite{sabatini2020, sabatini2024} for a full description of the observations and the data analysis.\\
\indent Considering the recent updates of dust temperatures for ATLASGAL clumps \citep{urquhart2022}, and following \cite{sabatini2020}, we derived the column densities using {\verb~MCWeeds~} \citep{giannetti2017} , which is an interface between {\verb~Weeds~} \citep{maret2011} and the Bayesian statistical models of {\verb~PyMC~} \citep{patil2010} , and we assumed $T_{\rm ex} = T_{\rm dust}$ equal to 20.3~K (G13.18+0.06), 15.8~K (G14.11-0.57), 19.1~K (G14.63-0.58), 12.1~K (G15.72-0.59), and 20.3~K (G19.88-0.54). The final column densities are in the range $\sim$($0.8$--$3.2)\times 10^{12}~$cm$^{-2}$ for o-$\mathrm{H_2D^+}$, $\sim$($3.7$--$5.9)\times 10^{11}~$cm$^{-2}$ for N$_2$D$^+$, $\sim$($0.2$--$3.0)\times 10^{13}~$cm$^{-2}$ for N$_2$H$^+$, $\sim$($0.6$--$4.0)\times 10^{13}~$cm$^{-2}$ for DCO$^+$, and $\sim$($1.6$--$6.3)\times 10^{13}~$cm$^{-2}$ for H$^{13}$CO$^+$. This matches the typical abundances relative to H$_2$ in high-mass star-forming clumps (e.g. \citealt{miettinen2020, sabatini2020, li2022}). The average error on the computed column densities is $\lesssim$20\%. We calculated $N$(HCO$^+$) from $N$(H$^{13}$CO$^+$) taking into account a $^{12}$C/$^{13}$C ratio that changes as a function of the galactocentric distance ($D_{\rm GC}$) of each source; that is, $[^{12}$C$]/[^{13}$C$] = 6.1^{+1.1}_{-1.8}\:D_{\rm GC}+14.3^{+7.7}_{-7.2}$ \citep{giannetti2014}. When applying Eq.~\eqref{eq:e_proxy} to this range of column densities, we obtained $N(\mathrm{e^-})$ in the range $\sim$($1.1$--$3.9)\pm0.5\times 10^{15}$cm$^{-2}$, corresponding to $X({\rm e}^{-})$ from 1.8$\times$10$^{-8}$ to 2.7$\times$10$^{-8}$. 
We note that a typical uncertainty of $\sim$30\% is associated with $X({\rm e}^{-})$, which results from propagating the uncertainty on the integrated line intensity. \\
\indent Regardless of the scale difference between observations ($\sim$1~pc) and simulation ($\sim$0.1~pc), HCO$^+$ is the dominating ion in both cases, together with H$_3^+$ (and its isotopologues) that is necessary to recover the electron abundance in the central region. This confirms the reliability of the method even when it is applied on different scales.
The simulations and observations we analysed clearly represent specific physical cases. We expect the formula to work also in situations in which other ions dominate the overall sum, in particular, cases in which small scales are properly resolved by high-resolution observations (e.g. using the Atacama Large Millimeter Array; ALMA) that might unveil emission and a contribution from more compact tracers.

\section{Conclusions}
\label{sec:conclusions}
We have introduced a CR propagation code for accurately estimating the CR ionisation rate in MHD simulations. We coupled our code to the outputs of MHD simulations performed with \textsc{gizmo} by implementing an interpolation scheme of gas properties consistent with the smoothing kernel procedure in the simulation. The propagator maps CRs as a flux of TPs for which the effective column density that is traversed following the magnetic field lines is determined by means of a Runge-Kutta integrator. The CR ionisation rate at the location of each TP was then estimated according to the $\mathscr{H}$ and $\mathscr{L}$ models by \citet{padovani2018} and was deposited onto the gas tracers in the original simulation in a self-consistent way. The results of the propagator were then used to investigate the evolution of several ions whose abundances are directly affected by CRs.

We tested the convergence of the code and found the optimal number of TPs (see Appendix A) that balance accuracy and performance (10240). We therefore demonstrated that this number is adequate enough to generate a reliable characterisation of the CR ionisation rate of the object because the differences from the most expensive run (32 times more TPs) are minimal and mostly affect the background region.

After the optimal number of TPs was found, we analysed the time evolution of the chemistry in a collapsing magnetised prestellar core from \citet{bovino2019} by employing the \textsc{krome} framework in post-processing as described by \citet{ferrada-chamorro2021}. We focussed on three different CR models, two variable cases ($\mathscr{H}$ and $\mathscr{L}$ in \citealt{padovani2018}) evolved by means of the propagation scheme, and a reference constant case with $\zeta_2=2.5\times 10^{-17}\rm\, s^{-1}$, with the aim to assess the CR propagation impact on the chemistry.

We found that a proper CR propagation scheme can produce significant differences (with a non-linear behaviour) for key chemical tracers, in particular, on deuterated species such as DCO$^+$ and N$_2$D$^+$.
We also analysed the correlations between the electron abundance $X({\rm e}^-)$ and $\zeta_2$ and between the electron column density and the column density of the most relevant positive ions. We found that clear correlations exist in both cases, and the two CR ionisation models $\mathscr{H}$ and $\mathscr{L}$ only yield a different normalisation. In particular, we provided an analytical function to estimate the electron abundance in star-forming regions with a standard deviation of 0.6, and we showed that when applied to observational data, it provides reliable estimates. For early stages high-mass clumps, we reported values around 10$^{-8}$ that agree with previous estimates of prestellar regions \citep[e.g.][]{caselli1998,bergin1999,maret2007}.
We analysed the correlation between $\mathscr{\zeta} _2$ and $n$(H$_2$) and determined the best-fit expressions, with an uncertainty of $\sim$25\% that could be employed as an approximate estimator for $\mathscr{\zeta} _2$ in numerical simulations without a self-consistent CR propagation scheme.

Finally, the increasing capabilities of modern astronomical facilities such as ALMA represent a viable method for testing our predictions. Additional high-sensitivity and high-resolution observations of the molecular tracers involved in Eq.~\eqref{eq:e_proxy} are needed to reveal possible fluctuations in electron abundance on core scales. In this context, intensive observational campaigns such as the recently accepted ALMA large program UNveiling the Initial Conditions of high-mass star-formation (UNIC)\footnote{\url{https://almascience.eso.org/alma-data/lp}}, which targets ten massive clumps with potentially more than 130 embedded cores, offer a unique opportunity to extend the results reported in this work under a variety of physical and chemical conditions in the near future.

\begin{acknowledgements}
        GL gratefully acknowledges the financial support of ANID-Subdirección de Capital Humano/Magíster Nacional/2024-22241873 and Millenium Nucleus NCN23{\_}002 (TITANs). SB acknowledges BASAL Centro de Astrofisica y Tecnologias Afines (CATA), project number AFB-17002. GS acknowledges the PRIN-MUR 2020 BEYOND-2p (Astrochemistry beyond the Second period elements, Prot. 2020AFB3FX), the project ASI-Astrobiologia 2023 MIGLIORA (Modeling Chemical Complexity, F83C23000800005), the INAF Mini-grant 2023 TRIESTE (“TRacing the chemIcal hEritage of our originS: from proTostars to planEts”; PI: G. Sabatini), and the National Recovery and Resilience Plan (NRRP), Mission 4, Component 2, Investment 1.1, Call for tender No. 104 published on 2.2.2022 by the Italian Ministry of University and Research (MUR), funded by the European Union – NextGenerationEU– Project Title 2022JC2Y93 Chemical Origins: linking the fossil composition of the Solar System with the chemistry of protoplanetary disks – CUP J53D23001600006 - Grant Assignment Decree No. 962 adopted on 30.06.2023 by the Italian Ministry of MUR.
\end{acknowledgements}

\bibpunct{(}{)}{;}{a}{}{,}
\bibliographystyle{aa}
\bibliography{refs}

\begin{appendix} 

\section{Convergence tests}
Because of our choice of kernel size, the scheme does not guarantee that all gas particles in the original simulation are used for the calculation. When this happens, $\zeta_2$ is set to zero, which would yield an inaccurate chemical evolution. In order to avoid this issue, there are two possible solutions. The computationally cheapest one, that however we have not considered in this work, is similar in spirit to the approach in \citet{ferrada-chamorro2021}, and interpolates $\zeta_2$ using the values of the neighbouring gas particles. The other possibility, that we have employed here, as it guarantees a much higher accuracy, requires a sufficiently large number of TPs, so that no regions of the cloud are left uncovered during the TP propagation. 

In order to identify the minimum number of particles required to reach convergence, that is our `fiducial' case, we explored different values of TPs in a suite of test runs, starting from 5120 TPs and gradually increasing their number by a factor of two, as reported in Table~\ref{tab:convergence}.

\begin{table}[h!]
    \caption{Number of TPs in the suite of convergence runs performed. The fiducial case is highlighted in boldface.}
    \begin{center}
    \begin{tabular}{ccccccc}
    \hline
    ID  & X1 & \textbf{X2} & X4 & X8 & X16 & X32 \\
    \#TP & 5120 & \textbf{10240} & 20480 & 40960 & 81920  & 163840\\
    \hline
    \end{tabular}
    \label{tab:convergence}
    \end{center}
\end{table}

\begin{figure}[h!]
\onecolumn
    \includegraphics[width=\textwidth]{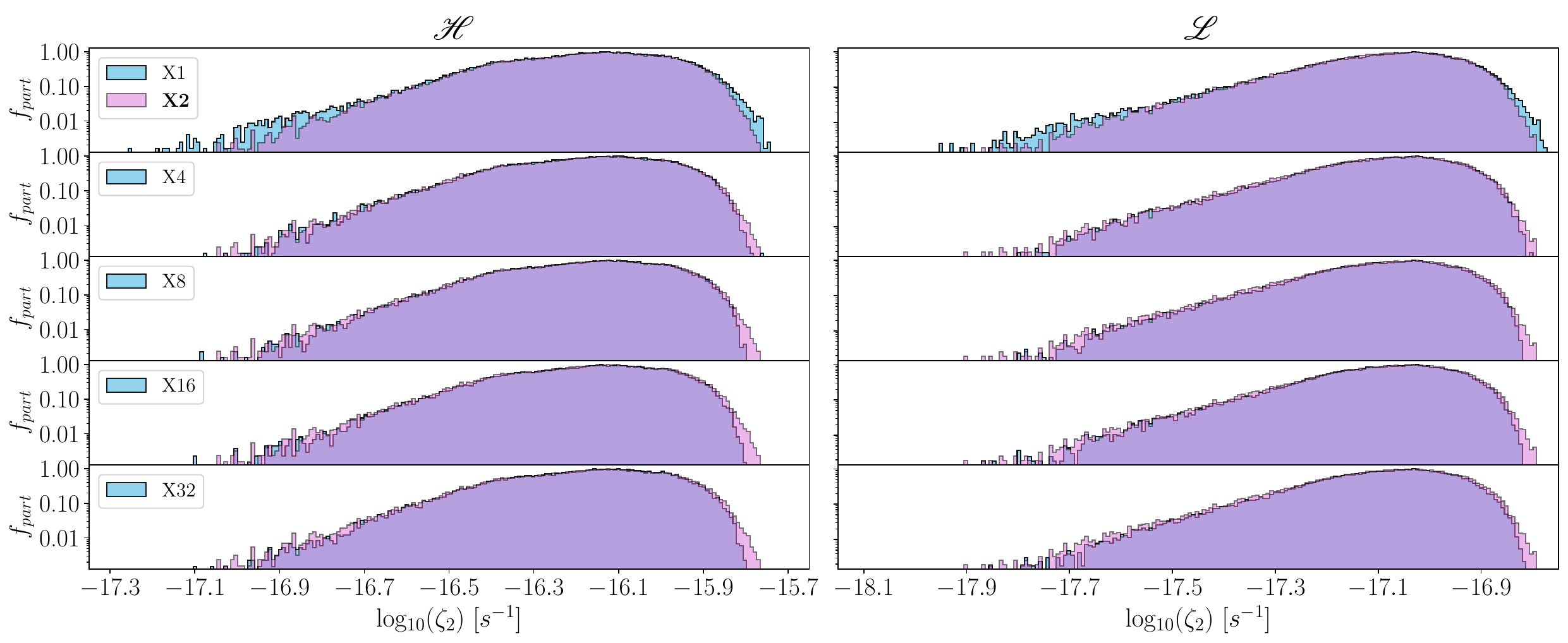}
    \caption{$\zeta_2$ distribution for $\mathscr{H}$ (left) and $\mathscr{L}$ (right) at 153.3~kyr in a 0.2$\times$0.2$\times$0.2~pc$^3$ box. In each row, we compared one of the runs in Table~\ref{tab:convergence} (in cyan) with our fiducial case (X2, in pink), with the matching regions in purple. $f_\mathrm{part}$ represents the fraction of particles.}
    \label{fig:conv_runs}
\twocolumn
\end{figure}
\newpage
In Fig.~\ref{fig:conv_runs}, we show the distribution of $\mathscr{\zeta}_2$ per particle for the $\mathscr{H}$ case (left-hand panels), and for the $\mathscr{L}$ case (right-hand panel), comparing each run in Table~\ref{tab:convergence} with our fiducial case (X2). Most of the particles are located on the rightmost side of the distribution, with a long tail extending to low $\zeta_2$ values. This reflects that gas particles with large $\zeta_2$ are located near the edge of the cloud where the attenuation is not strong yet, before the steep decrease begins as we move towards the centre of the cloud, which corresponds to the leftmost tail where the attenuation is the strongest. In general, we find a reasonable agreement in the shape of the distribution among the runs, which is consistent with the fact that the distribution is mainly determined by the physical properties of the cloud, which does not vary. The only differences we find can thus be attributed to the number of TPs used, which affects the smoothing procedure and the integration step. On one hand, X1 exhibits an overestimation at the edges of the distribution, because of the poor spatial sampling that is more sensitive to outliers. 
On the other hand, the increase of TPs slowly carves the high-$\zeta_2$ region, thanks to a better spatial sampling which removes the outliers, but does not significantly change the low-$\zeta_2$ tail, which instead shows stochastic variations due to the small number of gas particles in the central region. This comparison implies that the X2 case represents the best compromise between accuracy and efficiency for the chemical analysis we are interested in. For sake of completeness, in Appendix~\ref{app:rel_error} we analyse the relative error in the line-of-sight projection of $\zeta_2$ between our fiducial run and the other five runs.

\clearpage
\section{Relative error of expensive runs}
\label{app:rel_error}
\begin{figure}[h!]
\centering
    \includegraphics[width=\columnwidth]{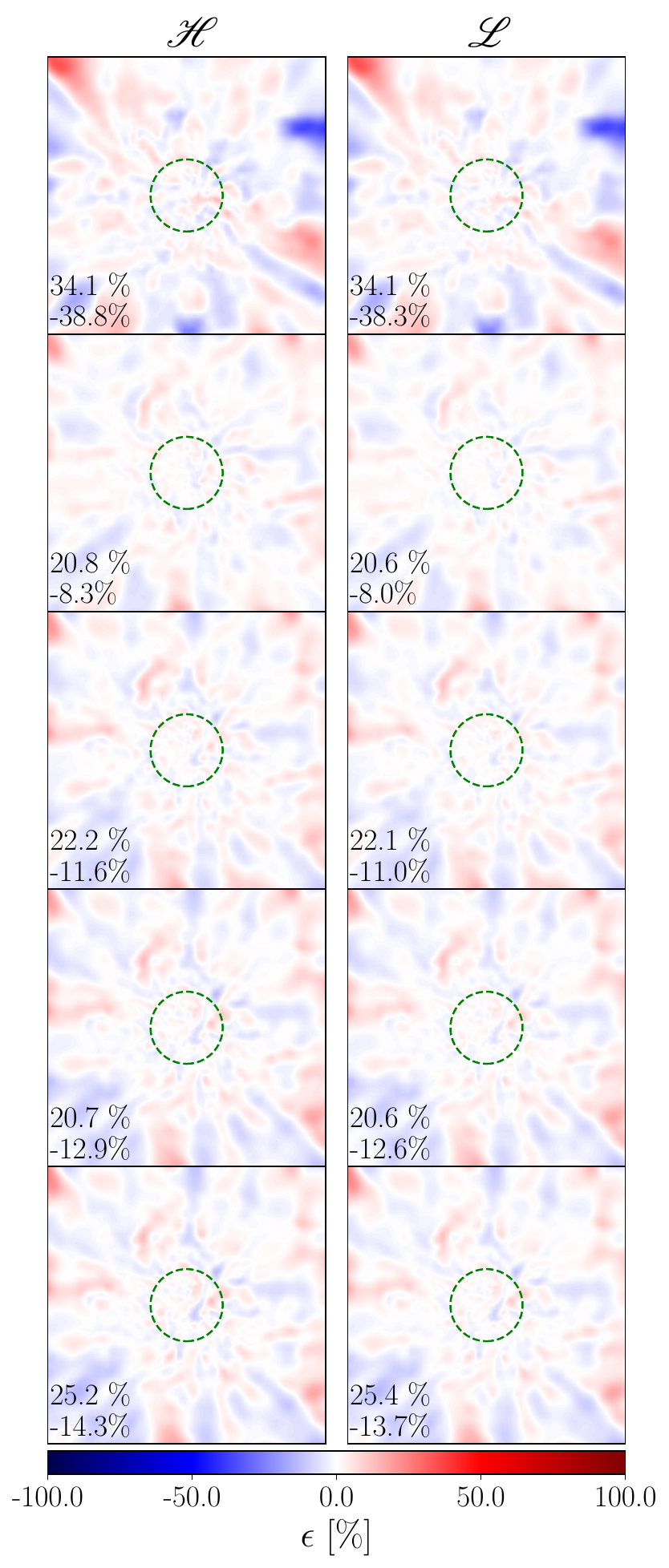}
    \caption{Relative error ($\epsilon$) for $\mathscr{H}$ (left) and $\mathscr{L}$ (right) within a 0.5x0.5x0.2~pc$^3$ at 153.3~kyr. We chose this box size to take into account the 0.5~pc of injection and a depth length similar to the filament scale (0.2~pc) to avoid the inclusion of background particles. The errors are calculated between X2 case and the rest. The panels are, from top to bottom: X1, X4, X8, X16 and X32. The bluer and redder indicate underestimation and overestimation of X2 relative to the others. The maximum and minimum errors for each comparison are written in the bottom-left corner. All green spheres have 0.065~pc radii.}
    \label{fig:rel_error}
\end{figure}

Fig.~\ref{fig:rel_error} shows the relative difference between $\mathscr{\zeta} _2$ obtained in our fiducial case, X2, and the other runs employing different numbers of TPs, averaged along a line-of-sight aligned with the z-axis. The largest and smallest relative error among the pixels are reported in the bottom left corner of every panel, corresponding to X1, X4, X8, X16, and X32 from top to bottom, respectively. We clearly observe larger deviations for the X1 case, whereas the other cases oscillate around 10-20\% almost independently of resolution.

To further check the robustness of our calculations, we determined the error within a spherical region of radius 0.065~pc (identified by the green dashed circle), comparable to the typical size of star-forming filaments. 
The relative error in this case drops by about a factor of 2, reaching at most $~3-13$\% for both ionisation models. The use of 10240 particles (our base) seems therefore adequate, as higher resolution runs do not significantly improve the results, but at the same time require a much higher computational cost. Obviously, on the largest scales explored here, the errors are larger, but this is expected as we start to encompass more background gas as well, which has lower column densities, hence is more easily affected by variations in $\zeta_2$.

\end{appendix}

\end{document}